\documentclass[11pt,reqno]{nature}

\usepackage[parfill]{parskip}  
\usepackage[english]{babel}
\usepackage{amsmath}
\usepackage{amsfonts}
\usepackage{amssymb}
\usepackage{amsthm}
\usepackage[pdftex]{graphicx}
\usepackage{epstopdf}
\usepackage{enumitem}
\usepackage{geometry}
\usepackage{hyperref}
\usepackage{tikz}
\usepackage[all]{xy}
\usepackage{multirow}
\usepackage{anysize}
\usepackage{color}
\usepackage{pdflscape}

\newcommand{\beq}{\begin{equation}}
\newcommand{\eeq}{\end{equation}}

\marginsize{3.2cm}{3.2cm}{3.2cm}{3.2cm}

\title{Do good actions inspire good actions in others?}

\author{Valerio Capraro$^1$ $\&$ Alessandra Marcelletti$^2$}

\begin{document}

\maketitle

\begin{affiliations}
\item Department of Mathematics, University of Southampton, SO17 1BJ, United Kingdom.
\item Department of Economics, University of Rome `Tor Vergata', 00133, Roma, Italy.
\end{affiliations}

\begin{abstract}
Actions such as sharing food and cooperating to reach a common goal have played a fundamental role in the evolution of human societies. These \emph{good} actions may not maximise the actor's payoff, but they maximise the other's payoff. Consequently, their existence is puzzling for evolutionary theories. Why should you make an effort to help others, even when no reward seems to be at stake? Indeed, experiments typically show that humans are heterogeneous: some may help others, while others may not. With the aim of favouring the emergence of `successful cultures', a number of studies has recently investigated what mechanisms promote the evolution of a particular good action. But still little is known about if and how good actions can spread from person to person. For instance, does being recipient of an altruistic act increase your probability of being cooperative with others? Plato's quote, `Good actions give strength to ourselves and inspire good actions in others', suggests that is possible. We have conducted an experiment on Amazon Mechanical Turk to test this mechanism using economic games. We have measured willingness to be cooperative through a standard Prisoner's dilemma and willingness to act altruistically using a binary Dictator game. In the baseline treatments, the endowments needed to play were given by the experimenters, as usual; in the control treatments, they came from a good action made by someone else. Across four different comparisons and a total of 572 subjects, we have never found a significant increase of cooperation or altruism when the endowment came from a good action. We conclude that good actions do not necessarily inspire good actions in others, at least in the ideal scenario of a lab experiment with anonymous subjects.
\end{abstract}

Humans are unique in the animal world for their willingness to help others and collaborate to reach a common goal and these attitudes are among the main reasons why human societies are so successful\cite{Tr,Ax-Ha,Os00,FF03,FF04,KG,No06,Ra-No,Ca}. Good actions, those which maximise the other's payoff, abound in the everyday life and experiments with anonymous people show that they are common even in the ideal setting of a lab, where confounding factors such as long-term strategies, indirect rewards, communication, signalling, etc., are not present\cite{CJR,CSMN,WH,A12,EZ,KL,BC}. 

Nevertheless, not everyone behaves in such good ways and clouds of selfish people can often be observed in both the everyday life and the lab. Some people perceive the individual cost of restricting their individual freedom of choice as being too large and decide to free-ride and aim for their personal benefit; consequently, people in connections with these free-riders typically decide to either break the link with the free-riders or to free-ride as well. Either way, this generates a cloud of defectors in the human social network\cite{RAC,WSW}.

To avoid this suboptimal scenario, scholars have started investigating what mechanisms can promote good actions in lab experiments. The underlying motivation is that, in case we knew that mechanism M promotes such behaviours, then institutions could use this mechanism to foster good actions to give rise to more successful societies.  

Here we consider three different kinds of good actions: altruism, benevolence, and cooperation. In the lab, altruistic attitudes are typically measured through the (binary) Dictator game (DG)\cite{KKT}, benevolent characteristics are measured using the Benevolence game (BG)\cite{CSMN}, and cooperative tendencies are measured through the Prisoner's dilemma (PD)\cite{Fl}. In the binary DG, one person, named dictator, is given an endowment of 100 Monetary Units (MUs), while the other person is given nothing. The dictator can either take all the 100MUs for himself and leave the other person with nothing or he can split the money evenly. The other person has no choice and receives what the dictator decides to donate. In the BG a person is given an endowment of 100MUs and has to decide between burning it or give it to another player. Also in this case, the other person has no say. In the PD, two people are given $100$MUs each and have to decide between hand it over or not. If a participant hands his money over, the other participant earns $100k$ MUs, where $k>1$. In the DG, it is optimal to keep all the money, but the good action is to split it; in the BG a money maximiser person is indifferent between his two strategies, while the good action is to let the other person get the endowment; in the PD, participants are better off not handing the money over, but the good action is to hand it over.

Virtually all studies have focused on whether mechanisms such as punishment of defectors\cite{BR,FG1,FG2,GIR}, reward of cooperators\cite{MSK,PB,MSKM} or a combination of these two\cite{AHV,RM,SSW,HS} can promote cooperative behaviour. Little is known about if and how good actions of possibly different nature can spread from person to person. For instance, does being recipient of an altruistic action increase your probability to cooperate with a third party? Plato's quote `Good actions give  strength to ourselves and inspire good actions in others' suggests that recipients of good actions should act in a better way with others than people who were recipient of a neutral action.  Besides the clear applications that this principle, if true, would have on finding ways to foster good behaviours in human societies, it may potentially have deep consequences also in economic theory: if our actions strongly depend on what others have previously done to us, the stability of our preferences over time would be seriously questioned. Perhaps we were selfish yesterday because someone was mean to us in the early morning, and we are altruist today because someone has been kind to us this morning. Reciprocity theories\cite{DK,FF} take this possibility into account when sequential interactions among the same agents are possible. However, Plato's principle is supposed to apply more generally, leading to the fundamental question: Does the behaviour of A towards B affect the behaviour of B towards C?

Here we start investigating this question by using economic games. Our typical experiment involves three people, A, B, and C. The description of the exact design is postponed to the Methods section. Abstractly speaking, all our experiments have the same basic structure: Person B is, at the same time, the target of a decision made by Person A, and the maker of a decision that can affect only Person C. We investigate whether the nature of A's action (good or neutral) affect the nature of B's action. Since we are interested in seeing how B's altruistic and cooperative tendencies change given the choice of A, Person B will either play a Dictator game or a Prisoner's dilemma. This gives rise to two baseline treatments, where B is asked to play either the DG or the PD, after being the recipient of a neutral action (endowment needed to play given by the experimenter). By varying the way the endowment is given to Person B by Person A (either an altruistic act in the DG or a benevolent act in the BG) we created four control treatments, where B is asked to play either the DG or the PD after being the recipient of a good action. In all four comparisons, and a total of 572 subjects, we have never found statistically significant difference in the behaviour of B towards C, depending on how A behaved towards B.

\section*{Method}

We recruited subjects using the online labour market Amazon Mechanical Turk (AMT)\cite{PCI,HRZ,R12}. As in classical lab experiments, AMT workers receive a baseline payment ($\$0.30$ in our case) and can earn an additional bonus depending on how they perform in the game. AMT experiments are easy to implement and cheap to realise, since AMT workers are paid a substantially smaller amount of money than people participating in physical lab experiments. Nevertheless, it has been shown that data gathered using AMT agree both qualitatively and quantitatively with those collected in physical labs in a variety of different strategical situations\cite{HRZ,R12,SW,RAC}. 

Yet there are some issues that may potentially invalidate data obtained using AMT. One of the major issues is that some subjects try to complete the HIT as fast as they can, to get the bonus in the shortest time, and so they may not fully understand the strategic situation that they are facing, increasing the risk of collecting meaningless data. We have addressed this problem in two different ways, depending on whether subjects acting as Player B had to play the DG or the PD. In the PD, we asked for four comprehension questions and we automatically screened out those subjects who failed any of the comprehension questions. Comprehension questions were formulated in order to make clear the tension between maximising one's own payoff and maximising the other's payoff. In the DG, since the strategical situation is really straightforward, we decided to skip the comprehension questions and ask the subjects to describe the reason of their choice. This procedure allowed us to check whether the subjects understood the decision problem, but also to address the second major issue of AMT experiments, namely that some subjects may think that the participants they are paired with are not real. Indeed, typically, in the description of the reason of their choice, one finds a few subjects saying `I kept the money because I think the other person is not real', or similar statements. Of course, we would like to exclude systematically this problematic subjects. To do so, we have manually checked all the original DataSet and we have constructed DataSetExcluded by removing those subjects belonging to one of the following categories: (i) Subjects who explicitly say that they believe that the other participant is not real; (ii) Subjects who do not provide any reason for their choice; (iii) Subjects whose reason of their choice was not consistent with their actual choice.  This led us with two datasets to analyse: DataSet and DataSetExcluded. Statistics are reported in the Results section, but we anticipate that the two analyses turned out to be qualitatively equivalent.

We now turn back to our research question and describe our two studies in more details (full instructions are reported in the Appendix). In Study 1 we have investigated whether recipients of a good action behave more cooperatively than recipients of a neutral action. In the baseline treatment, subjects in the role of Person B participated in a neutrally framed one-shot PD. Each participant was given an endowment of $\$0.20$ and paired with another anonymous participant (Person C). They could either keep the $\$0.20$ or hand it over. In this latter case, the other participant would earn $\$0.40$. As mentioned before, the decision was made after passing the comprehension questions. The two control treatments differed from the baseline treatment only in the way the initial $\$0.20$ was given to the participants. In Control 1, participants were informed that they had been previously paired with another participant (Person A), different from Person C, who was the dictator in a DG. Person A was given $\$0.40$ and could decide between keeping it all or splitting it with Person B and they decided to split it. Thus, in Control 1, the endowment needed to play the PD comes from an altruistic act of Person A. Control 2 was similar to Control 1, with the only difference that the $\$0.20$ comes from a BG. Specifically, participants were informed that they had been previously paired with another anonymous participants (Person A) who had to decide between doing nothing or making the benevolent act of letting the other player getting $\$0.20$ at zero cost to themselves. After making their decision, in each treatment participants entered the demographic questionnaire, where we asked for their gender, age, level of education and, finally, the reason of their choice in the game.

In Study 2 we have tested whether recipients of a good action behave more altruistically than recipients of a neutral action. In the baseline treatment, subjects participated to a binary DG. Each participant was given an endowment of $\$0.20$ and had to decide between keeping it and splitting it evenly with the other participant. Also in this case, the two control treatments differed from the baseline only in the way the initial $\$0.20$ was provided. In Control 1 they came from a donation in a previous DG; in Control 2 they came from a benevolent act in a previous BG. As mentioned before, there are no comprehension questions in Study 2, but, as in Study 1, after making their decision, subjects enter the demographic questionnaire, where we asked for their gender, age, level of education, and, importantly, the reason of their choice in the game.

After collecting the results, subjects were matched and bonuses were computed and paid. No deception was used. Informed consent was obtained by all participants. These experiments were approved by the Southampton University Ethics Committee on the Use of Human Subjects in Research.

\section*{Results} 

A total of 232 US subjects ($62.9\%$ male, mean age = 30.4) participated in our Study 1, as Person B, passing all comprehension questions. 81 subjects ($64.2\%$ male, mean age = 31.1) participated in the baseline and played a neutrally framed PD using an endowment provided by the experimenter; 75 subjects ($58.7\%$ male, mean age = 30.4) participated in Control 1 and played the PD using an endowment coming from a donation in a previous DG; 76 subjects ($65.8\%$ male, mean age = 29.7) participated in Control 2 and played the PD using an endowment coming from a benevolent act in a previous BG. The average cooperation is very similar across the three treatments ($33.3\%$ in the baseline, $33.3\%$ in Control 1, $25.0\%$ in Control 2). To test for an effect of how the endowment is provided (baseline vs Control 1 and baseline vs Control 2) we use logistic regression, with and without control on gender, age, and level of education, predicting cooperation or defection as the dependent variable. As shown by Table \ref{ta:PD} we find no significant effect of how the endowment was provided. Table \ref{ta:DGall} shows the effect of demographics on cooperation: none of the demographic characteristics we collected predicts cooperation significantly. As mentioned in the Methods section, we now report the statistical analysis of DataSetExcluded. In a manual screening, we excluded 9 subjects and remained with 223 US subjects ($62.8\%$ male, mean age = 30.5). 79 subjects ($62.9\%$ male, mean age = 31.3) in the baseline, 71 subjects ($60.6\%$ male, mean age = 30.4) in Control 1, and 74 subjects ($64.9\%$ male, mean age = 29.8) in Control 2. Also in this case, the average cooperation was very similar across the three treatments ($33.3\%$ in the baseline, $32.3\%$ in Control 1, and $25.7\%$ in Control 2) and Table 1 shows that the way the endowment was provided had no statistically significant effect on cooperative behaviour.  We conclude, that a good action in the DG or in the BG does not inspire a good action in the PD, at least in our subject pool. Impressively, reading through all reasons provided by the subjects in the control treatments, we discovered that literally nobody declared that his or her action in the PD was somehow influenced by how they were previously treated.

A total of 340 US subjects ($63.9\%$ male, mean age = 39.2) participated in our Study 2, as Person B. 112 subjects ($59.8\%$ male, mean age = 28.1) participated in the baseline and played a neutrally framed binary DG with endowment provided by the experimenter; 115 subjects ($57.4\%$ male, mean age = 28.7) participated in Control 1 and played the binary DG with endowment coming from a donation in a previous binary DG; 113 subjects ($74.4\%$ male, mean age = 30.7) participated in Control 2 and played the binary DG with endowment coming from a benevolent act in a previous BG. The average choice (Keep = 0, Split =1) is very similar across the three treatments ($0.43$ in the baseline, $0.42$ in Control 1, $0.48$ in Control 2). To test for an effect of how the endowment is provided (baseline vs Control 1 and baseline vs Control 2) we use logistic regression, with and without control on gender, age, and level of education, predicting keeping or splitting as the dependent variable. As shown by Table \ref{ta:DG} we find no significant effect of how the endowment was provided. Table \ref{ta:DGall} shows the effect of demographics on cooperation: females donated significantly more than males and age had a borderline significant positive effect on altruism. As mentioned in the Methods section, we now report the statistical analysis of DataSetExcluded. In a manual screening, we excluded 24 subjects and remained with 316 US subjects ($62.7\%$ male, mean age = 29.3). 108 subjects ($58.3\%$ male, mean age = 28.1) in the baseline, 102 subjects ($54.9\%$ male, mean age = 29.3) in Control 1, and 106 subjects ($74.5\%$ male, mean age = 30.5) in Control 2. Also in this case, the average choice was very similar across the three treatments ($0.45$ in the baseline, $0.47$ in Control 1, and $0.50$ in Control 2) and Table 1 shows that the way the endowment was provided had no statistically significant effect. We conclude, that a good action in the DG or in the BG does not inspire a good action in the DG, at least in our subject pool. We mention that, this time, reading through all reasons provided by the participants in the control treatments shows that several players were influenced by how they were previously treated. To be more precise, 15 out of 102 subjects in Control 1 Excluded declared `The other participant chose to split his 40 cents with me, so I elected to pass on the love and do the same' or equivalent statements. A similar thing happened in Control 2. Thus there is a psychological effect, but it does not give rise to an economically and statistically significant effect. It is then possible that those people who reported to be positively influenced by the other's choice would split the money anyway. 

Finally, a total of 175 US subjects were recruited to play the role of Person A (90 in the DG and 85 in the BG). Note that the statistics in Table \ref{ta:DGall} do not include these participants, since they played a different DG ($\$0.40$ at stake, instead of $\$0.20$). Statistics on these subjects are not significant, probably due to the relatively small sample. We used these subjects only to avoid deception and match the players in the role of Person B with a real participant.

\section*{Discussion}

Good actions are defined as those which maximise the other's payoff. Sharing the endowment in the binary Dictator game (DG), letting the other person take the endowment in the Benevolence game (BG), cooperating in the Prisoner's dilemma (PD), are all examples of good actions. Motivated by Plato's quote `Good actions give strength to ourselves and inspire good actions in others' we have investigated whether good actions of possible different types can spread from person to person in the simplest possible way: does a good action of A towards B increase the probability of a good action of B towards C? We have conducted six experiments: two baseline treatments where B has to make a decision in either the DG or the PD using an endowment given by the experimenter; four control treatments where the endowment needed to play comes from either an altruistic action or a benevolent action made by someone else in a previous interaction. Across four comparisons and 572 participants we have never found a significant increase of good actions by player B when they were recipient of a good action.

Our results provide evidence that good actions do not spread from person to person, at least in our ideal case where interactions happen through Amazon Mechanical Turk and are therefore completely anonymous. 

However, we recommend caution in the interpretation of our results. One of the major limitations of AMT experiments is that it is virtually impossible to convince the participants that the people they interact with are real. To address this problem, we asked the subjects to write the reason of their choice and conducted two statistical analyses, one including all participants and one including only those whose description revealed that they were playing as they would do in reality. Although the two analyses gave qualitatively equivalent results, we cannot be completely sure that \emph{all} participants in the second analysis acted as they would act in a real scenario: it is possible that they described a reason consistent with a real scenario, but behaved as the other participants were not real. On the other hand, it is extremely difficult to design an experiment where participants are made completely convinced that the other participants are real, without generating confounding factors. For instance, showing pictures, names, or TurkIDs would decrease the `distance' between the participants with the consequence of increasing good actions, as shown by a number of similar studies\cite{FOM,B03,CG08}. Consequently, it would be difficult to understand which cause gives rise to which effect. This issue connects to the problem of whether good actions inspire good actions in others in the everyday life. Everyday good actions are typically accompanied by eye-contact, small talks, smiles, and other kinds of signals. It is likely that these signals contribute to improve the recipient's mood so that the recipient's utility is the sum of his material payoff and his nonmaterial payoff given by the fact that his mood has been indirectly improved by other factors. It is then possible that it is not really the fact of being recipient of a good action that inspires other good actions, but the fact that the mood has been improved. 

Our results also contribute to the increasing body of literature regarding gender differences in the Dictator game. Alike the majority of studies\cite{EG,AV,DM,HS09,KC,D14,B14,DEJR}, but not all\cite{BK95,B14,DEJR}, we too have found that females are significantly more generous than males in the Dictator game. Furthermore, consistent with Engel's meta-study\cite{En}, we too have found that age has a positive effect on giving in the DG, though our effect is only borderline significant ($P=0.05$ in DG and $P=0.092$ in DG excluded).

Finally, our results add to the research concerning framing effects in the Dictator game and the Prisoner's dilemma. A number of studies\cite{KC,Su,DEJR} agree that behaviour in the Dictator game is independent of the name of the game (Keep game vs Take game) and the name of the strategies (Keep vs Give). Here we have shown that it is independent of how the endowment is provided (by the experimenter vs by someone else through a good act). Similarly, a number of studies\cite{LSR,KR,A95,BK,Co,EJMM,MS,MC,PC,SS,vDK,ZLM}, but not all\cite{CDG,DGH,RT}, suggest that behaviour in the Prisoner's dilemma depends on the name of the game and the name of the strategies. Here we have shown that it does not depend on how the endowment is provided.

%\section*{Methods Summary}

%We recruited US subjects using Amazon Mechanical Turk and randomly assigned them to one of eight experiments using economic games. Treatments are described in the Main Text and full instructions are reported in the Supporting Information. We asked four comprehension questions to make sure that participants understood the rules of the games. Participants failing any of the comprehension questions were automatically screened out and could not participate to the experiment. Participants were informed about this procedure at the very beginning of the experiment and could decide to opt out at that stage as well as at every stage thereafter. A total of 858 subjects passed all comprehension questions and participated in our experiments, earning $\$0.30$ as a participation fee. Participants were also informed that computation and payment of the bonuses would be made at the end of the experiment. So, importantly, in each treatment, participants played the second game without knowing the outcome of the first. No deception was used. Informed consent was obtained by all participants. These experiments were approved by the Southampton University Ethics Committee on the Use of Human Subjects in Research. 

\begin{addendum}
 %\item We thank Elena Simperl for useful help on the design of the experiment and David G. Rand for comments on the first draft of this paper.
 \item[Competing Interests] The authors declare that they have no competing financial interests.
 \item[Correspondence] Correspondence and requests for materials should be addressed to V.C.~(email: v.capraro@soton.ac.uk).
\end{addendum}

%%
%% TABLES
%%
%% If there are any tables, put them here.
%%

\section*{Full Instructions of Study 1}

Here we report the full instructions for Control 1. The instructions for Baseline and Control 2 were very similar and we highlight the differences along the description. 

The first two screens do not contain any information about the game and serve us only as control to avoid multiple plays from the same subject and lazy participants who can increase randomness on our data.

\textbf{Screen 1.} In the first screen, participants were welcomed to the game and asked to type their worker ID. This allows us to automatically exclude workers who have already completed the task.

\textbf{Screen 2.} In the second screen, we asked the participant to transcribe a relatively long neutral piece of text. This allows us as to tell computers and humans apart (CAPTCHA) and, at the same time, to exclude lazy workers and minimise randomness in our data. We used a meaningless neutral text in order to avoid framing effects.

In the third screen, people entered the real game. Here is the exact instructions we used.

\textbf{Screen 3.} Welcome to this HIT. This HIT will take about ten minutes. For the participation to this HIT, you will earn $\$0.30$. You can also earn additional money depending on the decision you and the other participants will make. You will be asked to make two decisions. There is no incorrect answer. However, to make sure you understand the situation, we will ask some simple questions, each of which has only one correct answer.  If you fail to correctly answer any of those questions, the survey will automatically end and you will not receive any redemption code and consequently you will not get any payment. With this in mind, do you wish to continue? 

Here participants could either continue or end the survey, clicking on the corresponding button. Participants who decided to continue were directed to the next screen. 

\textbf{Screen 4 (Control 1)} In a previous part of the HIT, which you have not seen, you were paired with another participant. This participant was given $\$0.40$ and had to decide between keeping it all or splitting it evenly with you. He decided to split. So.. congratulations! You now have $\$0.20$.

\textbf{Screen 4 (Control 2)} In a previous part of the HIT, which you have not seen, you were paired with another participant. The other participant was told that there were $\$0.20$ available but she or he could not get it. They could only choose between doing nothing or donating it to you. He decided to donate. So.. congratulations! You now have $\$0.20$.

\textbf{Screen 4 (Baseline)} In the Baseline treatment participants jumped from Screen 3 to Screen 5.

\textbf{Screen 5.} You have been paired with another participant, different from the one you were paired before. The amount of money you can earn depends on your and the other participant's decision. You and the other participant have both $\$0.20$, earned in the previous part of the HIT. You must decide whether to hand it over or not. Each time a participant hands over their $\$0.20$, the other participant earns $\$0.40$. So:
\begin{enumerate}
\item If you both decide to hand over the $\$0.20$, you end the game with $\$0.40$
\item If the other participant hands it over and you do not, you end the game with $\$0.60$
\item If you hand it over and the other participant does not, you end the game with $\$0$ 
\item If neither of you hand it over, then you end the game with $\$0.20$
\end{enumerate}

Of course, in the Baseline, we took out the sentence `different from the one you were paired before'.

\textbf{Screen 6.} Here are some questions to ascertain that you understand the rules. Remember that you have to answer all of these questions correctly in order to get the completion code. If you fail any of them, the survey will automatically end and you will not get any payment. 
\begin{enumerate}
\item What choice should YOU make to maximise YOUR gain?
\item What choice should YOU make to maximise the OTHER PARTICIPANT's gain?
\item What choice should the OTHER PARTICIPANT make to maximise THEIR gain?
\item What choice should the OTHER PARTICIPANT make to maximise YOUR gain?
\end{enumerate}

In each questions, participants could answer by selecting either `Do not hand over' or `Hand over'. Participants failing any of the comprehension questions were automatically screened out through a `Skip Logic', which is very easy to implement using the survey builder Qualtrics.

\textbf{Screen 7.} Now it's time to make your decision. What is your choice?

Here participants could select either `Do not hand over' or `Hand over'. Following this screen, we asked demographic questions and the description of the reason of their choice. A final screen, providing a completion code to claim for their payment, concluded the survey.

\section*{Full Instructions of Study 2} 

Here we report the full instructions for Control 1. The instructions for Baseline and Control 2 were very similar and we highlight the differences along the description. Moreover, since some screens were identical to those of Control 1 in Study 1, we report only those screen containing some differences.

\textbf{Screen 5.} You have been paired with another participant, different from the one you were paired before. You must decide between keeping all your $\$0.20$ or splitting it evenly with the other participant. This decision is unilateral. The other participant does not have the possibility to influence your payoff.

Of course, in the Baseline, we took out the sentence `different from the one you were paired before'.

\textbf{Screen 6.} As mentioned in the Main Text, Study 2 does not contain any comprehension questions. So subjects passed directly from Screen 5 to Screen 7.

\textbf{Screen 7.} What is your choice?

Here participants could select either `Keep' or `Split'. Following this screen, we asked demographic questions and the description of the reason of their choice. A final screen, providing a completion code to claim for their payment, concluded the survey.

\begin{landscape}
\begin{table}\caption{The effect of being recipient of a good action on cooperation in the Prisoner's Dilemma. We used logistic regression with and without control on sex, age, and education, using `treat' as a dummy variable. We report the $\beta$-value, its standard error, and the significance level. `Bas' stands for `Baseline', `Con' for `Control', and `Exc' for `Excluded. So, for instance, the column `Bas Exc vs Con 1 Exc' reports the results of the regression using Bas Exc = 0 and Con 1 Exc =1 as dummy variable. We find no significant effect of the dummy variable on cooperation in the PD. Being recipient of a good action does not significantly increase the probability of cooperating in the PD. \label{ta:PD}}
\begin{center}
\begin{tabular}{l*{8}{rr}}
\hline
                &\multicolumn{1}{c}{Bas vs}&\multicolumn{1}{c}{Bas vs}&\multicolumn{1}{c}{Bas Exc vs}&\multicolumn{1}{c}{Bas Exc vs}&\multicolumn{1}{c}{Bas vs}&\multicolumn{1}{c}{Bas vs}&\multicolumn{1}{c}{Bas Exc vs}&\multicolumn{1}{c}{Bas Exc vs}\\
                &         Con 1  &         Con 1   &         Con 1 Exc   &        Con 1 Exc   &         Con 2   &         Con 2   &         Con 2 Exc    &        Con 2 Exc   \\
\hline
dummy for decision&                &                &                &                &                &                &              
   &                \\
treat           &       --.000   &         .043   &       --.405   &       --.419   &       --.405   &       --.419   &       --.405   
 &       --.419   \\
                &        (.34)   &        (.34)   &        (.36)   &        (.36)   &        (.36)   &        (.36)   &        (.36)   
 &        (.36)   \\
sex             &                &         .020   &                &       --.198   &                &       --.198   &                
 &       --.198   \\
                &                &        (.35)   &                &        (.38)   &                &        (.38)   &                
 &        (.38)   \\
age             &                &       --.012   &                &       --.004   &                &       --.004   &                
 &       --.004   \\
                &                &        (.02)   &                &        (.02)   &                &        (.02)   &                
 &        (.02)   \\
education       &                &         .291   &                &         .252   &                &         .252   &                
 &         .252   \\
                &                &        (.16)   &                &        (.17)   &                &        (.17)   &                
 &        (.17)   \\
Constant        &       --.693** &      --1.620   &       --.693** &      --1.399   &       --.693** &      --1.399   &       --.693** 
 &      --1.399   \\
                &        (.24)   &        (.99)   &        (.24)   &       (1.02)   &        (.24)   &       (1.02)   &        (.24)   
 &       (1.02)   \\
\hline
Pseudo $R^2$    &         .000   &         .019   &         .007   &         .019   &         .007   &         .019   &         .007   
 &         .019   \\
No. of cases    &          156   &          156   &          157   &          157   &          157   &          157   &          157   
 &          157   \\
\hline

\end{tabular}
\end{center}
\begin{footnotesize}
\begin{flushleft}
\textbf{Significance level}:\hspace{1em} *** : $p<0.001$  \hspace{1em} ** : $p<0.01$
\hspace{1em} * : $p<0.05$ \\
%\emph{Notes}: Dependent variable: It is equal to to 0 if dgdec=1 and equal to 1 if dgdec=2. Sex is a dummy variable equal to 1 if XXX.\\
\end{flushleft}
\end{footnotesize}

\end{table}
\end{landscape}

\newpage
\begin{landscape}
\begin{table}\caption{The effect of being recipient of a good action on altruism in the Dictator game. We used logistic regression with and without control on sex, age, and education, using `treat' as a dummy variable. We report the $\beta$-value, its standard error, and the significance level. `Bas' stands for `Baseline', `Con' for `Control', and `Exc' for `Excluded. So, for instance, the column `Bas Exc vs Con 1 Exc' reports the results of the regression using Bas Exc = 0 and Con 1 Exc =1 as dummy variable. We find no significant effect of the dummy on donations in the DG. Being recipient of a good action does not significantly increase the probability of splitting the endowment in the DG.\label{ta:DG}}
\begin{center}
\begin{tabular}{l*{8}{rr}}
\hline
                &\multicolumn{1}{c}{Bas vs}&\multicolumn{1}{c}{Bas vs}&\multicolumn{1}{c}{Bas Exc vs}&\multicolumn{1}{c}{Bas Exc vs}&\multicolumn{1}{c}{Bas vs}&\multicolumn{1}{c}{Bas vs}&\multicolumn{1}{c}{Bas Exc vs}&\multicolumn{1}{c}{Bas Exc vs}\\
                &         Con 1  &         Con 1   &         Con 1 Exc   &        Con 1 Exc   &         Con 2   &         Con 2   &         Con 2 Exc    &        Con 2 Exc   \\
\hline

dec             &                &                &                &                &    
             &                &                &                \\
treat           &       --.082   &       --.115   &         .068   &         .024   &    
      .163   &         .257   &         .186   &         .289   \\
                &        (.27)   &        (.27)   &        (.28)   &        (.28)   &    
     (.27)   &        (.28)   &        (.27)   &        (.29)   \\
sex             &                &         .545   &                &         .436   &    
             &         .624*  &                &         .619*  \\
                &                &        (.28)   &                &        (.29)   &    
             &        (.31)   &                &        (.31)   \\
age             &                &         .030   &                &         .028   &    
             &         .015   &                &         .015   \\
                &                &        (.02)   &                &        (.02)   &    
             &        (.02)   &                &        (.02)   \\
education       &                &         .047   &                &       --.006   &    
             &       --.194   &                &       --.229   \\
                &                &        (.11)   &                &        (.12)   &    
             &        (.11)   &                &        (.12)   \\
Constant        &       --.251   &      --2.078** &       --.186   &      --1.556*  &    
    --.251   &       --.746   &       --.186   &       --.520   \\
                &        (.19)   &        (.74)   &        (.19)   &        (.76)   &    
     (.19)   &        (.74)   &        (.19)   &        (.77)   \\
\hline
Pseudo $R^2$    &         .000   &         .032   &         .000   &         .023   &         .001   &         .032   &         .002    &         .035   \\

No. of cases    &          227   &          227   &          210   &          210   &    
       225   &          225   &          214   &          214   \\
\hline
\end{tabular}
\end{center}
\begin{footnotesize}
\begin{flushleft}
\textbf{Significance level}:\hspace{1em} *** : $p<0.001$  \hspace{1em} ** : $p<0.01$
\hspace{1em} * : $p<0.05$ \\
%\emph{Notes}: Dependent variable: It is equal to to 0 if dgdec=1 and equal to 1 if dgdec=2. Sex is a dummy variable equal to 1 if XXX.\\
\end{flushleft}
\end{footnotesize}
\end{table}
\end{landscape}

%\begin{landscape}
\begin{table}\caption{Impact of demographic variables on individual decision in the DG and the PD before and after exclusion of subjects. We used logistic regression and report the $\beta$-value, its standard error, and its significance level. Females donate significantly more than males in the DG. Borderline positive effects of age on altruism in the DG and education on cooperation in the PD were also noted.\label{ta:DGall}}
\begin{center}
\begin{tabular}{l*{4}{rr}}
\hline
                &\multicolumn{1}{c}{DG}&\multicolumn{1}{c}{DG Exc}&\multicolumn{1}{c}{PD}&\multicolumn{1}{c}{PD Exc}\\
             %   &         b/se   &         b/se   &         b/se   &         b/se   \\
\hline
dec             &                &                &                &                \\
sex             &         .671***&         .594** &         .136   &         .060   \\
                &        (.18)   &        (.19)   &        (.30)   &        (.30)   \\
age             &         .015*   &         .017   &       --.001   &       --.004   \\
                &        (.01)   &        (.01)   &        (.01)   &        (.01)   \\
education       &       --.092   &       --.150   &         .209   &         .278*  \\
                &        (.08)   &        (.08)   &        (.14)   &        (.14)   \\
Constant        &      --1.140*  &       --.711   &      --1.886*  &      --1.994*  \\
                &        (.46)   &        (.48)   &        (.81)   &        (.84)   \\
\hline
Pseudo $R^2$    &         .029   &         .029   &         .010   &         .015   \\
No. of cases    &          340   &          316   &          232   &          223   \\
\hline

\end{tabular}
\end{center}
\begin{footnotesize}
\begin{flushleft}
\textbf{Significance level}:\hspace{1em} ***: $p<0.001$  \hspace{1em} ** : $p<0.01$
\hspace{1em} *$ : p<0.05$ \\
%\emph{Notes}: Dependent variable: It is equal to 0 if dgdec=1 and equal to 1 if dgdec=2. Sex is a dummy variable equal to 1 if XXX.\\
\end{flushleft}
\end{footnotesize}
\end{table}
%\end{landscape}

\end{document}